\documentclass[acmsmall,screen,nonacm]{acmart}

\usepackage{url}
\usepackage{listings, lstautogobble}
\usepackage{xspace}
\usepackage{graphicx}
\usepackage[most]{tcolorbox}
\usepackage{xcolor}
\usepackage{tikz}
\usetikzlibrary{arrows.meta,calc}

\definecolor{myhighlight}{RGB}{255,255,102} % A light yellow color

\usepackage{wrapfig}

\usepackage[nameinlink]{cleveref}
\usepackage{hyperref}

\newcommand{\projName}{\textsc{MockMaster}\xspace}
\newcommand{\tecton}{\textsc{Tecton}\xspace}

\newcommand{\eg}{\hbox{\emph{e.g.}}\xspace}
\newcommand{\ie}{\hbox{\emph{i.e.}}\xspace}
\newcommand{\etc}{\hbox{\emph{etc.}}\xspace}
\newcommand{\vs}{\hbox{\emph{vs.}}\xspace}

\newcommand{\etal}{\hbox{\emph{et al.}}\xspace}
\newcommand{\wrt}{\hbox{\emph{w.r.t.}}\xspace}

\newcommand{\totalbenchmarks}{5\xspace}

\newboolean{showcomments}
\setboolean{showcomments}{true} % comment this line to deactivate comments
\ifthenelse{\boolean{showcomments}}{
  \newcommand{\nbc}[3]{
    {\colorbox{#3}{\bfseries\sffamily\scriptsize\textcolor{white}{#1}}}%
    {\textcolor{#3}{\textsf\small$\blacktriangleright$\textit{#2}$\blacktriangleleft$}}}
  \newcommand{\todo}[1]{\nbc{TODO}{#1}{red}\xspace}
}{
  \newcommand{\nbc}[3]{}
  \renewcommand{\todo}[1]{}
}

\newcommand{\cf}[1]{\texttt{#1}}

\definecolor{cgreen}{rgb}{0.25,0.5,0.35} % comments
\definecolor{stringred}{rgb}{0.6,0,0} % for strings

\lstdefinelanguage{bosque}{
keywords={function, op, api, model, scalar, extends, enum, type, of, requires, ensures, softchk, var, let, return, if, elif, else, assert, invariant, field, pred, fn, entity, env, resources, datatype},
keywordstyle=\color{blue}\bfseries,
identifierstyle=\color{black},
alsoother={@},
sensitive=true,
comment=[l]{\%\%},
commentstyle=\color{cgreen}\bfseries,
string=[b]{"},
stringstyle=\color{stringred}\ttfamily,
}
 
\usepackage{algorithm}
\usepackage{algpseudocode}

\setcopyright{acmlicensed}
\copyrightyear{2025}
\acmYear{2025}
\acmDOI{XXXXXXX.XXXXXXX}

\begin{document}

\title{\projName: How to Automock Your Friends and Services}

\author{S M Sadrul Islam Asif}
\email{sadrul.islam@uky.edu}
\affiliation{%
  \institution{University of Kentucky}
  \city{Lexington}
  \state{Kentucky}
  \country{USA}
}

\author{Kennett Puerto}
\email{kennett.puerto@uky.edu}
\affiliation{%
  \institution{University of Kentucky}
  \city{Lexington}
  \state{Kentucky}
  \country{USA}
}

\author{Earl T. Barr}
\email{e.barr@ucl.ac.uk}
\affiliation{%
  \institution{University College London}
  \city{London}
  \country{UK}
}

\author{Mark Marron}
\email{marron@cs.uky.edu}
%\orcid{0000-0003-3589-4860}
\affiliation{%
  \institution{University of Kentucky}
  \city{Lexington}
  \state{Kentucky}
  \country{USA}
}

\begin{abstract}
	%TeX root = mockmaster.tex

Modern cloud applications are built from independent components that communicate over APIs. Testing these applications is challenging because 
their components may be difficult to deploy, be owned by other teams, or have test-disrupting side-effects. Developers therefore turn to mocks 
to simulate such components. 
Manually written mocks are, however, brittle, costly to maintain, prone to errors, and typically cover only a small fraction of possible behaviors; relying on them confronts developers with 
the testing version of the infamous regular-expression adage: 
\emph{``I had a testing problem, so I wrote a mock. Now I have two problems.''}

We present \projName, a system that automatically generates mocks for cloud APIs and service-based applications 
from their API specifications. \projName combines LLM agents with symbolic techniques, eliminating the 
need for developers to implement mocks by hand. \projName can repeatedly execute a test while varying mock behavior, including failures and 
edge cases. Across $10$ RESTful service benchmarks comprising $65$kloc, \projName achieves industry standard code coverage of excellent/acceptable
for $9$ of the $10$, discovers a total of $119$ distinct 500 errors, and eliminates flaky tests, all while costing less than $\$0.20$ per service.

\end{abstract}

%\begin{CCSXML}
%<ccs2012>
%  <concept_id>00000000.0000000.0000000</concept_id>
%  <concept_desc>Do Not Use This Code, Generate the Correct Terms for Your Paper</concept_desc>
%  <concept_significance>500</concept_significance>
%</ccs2012>
%\end{CCSXML}

%\ccsdesc[500]{Do Not Use This Code~Generate the Correct Terms for Your Paper}

%\keywords{Keywords!!!}

\maketitle

\section{Introduction}
\label{sec:intro}
%TeX root = mockmaster.tex

Modern cloud applications are built from multiple components that communicate over APIs. Directly testing such applications can be difficult or infeasible: components may be owned by other teams, difficult to deploy in a test environment, or have side-effects that make repeated testing impractical. To overcome these problems, developers commonly use mocks to simulate these components during testing. Manually written mocks are no panecea: they are brittle, error-prone, and costly to maintain~\cite{mockingstudy,mockingstudy2}. Worse, developers must decide which behaviours a mock will exhibit and implement each of them. Minimising effort encourages mocks that typically cover only a small subset of behaviours, concentrating on common or happy-path responses. 

We address this limitation by treating mock generation as a sampling problem. The space of possible responses from even a single API can be enormous, making exhaustive testing impractical. Independently sampling complete responses can repeatedly exercise similar behaviours while leaving other parts of the response space unexplored. \projName{} instead uses the API specification to decompose a response into its constituent components and an LLM to generate diverse, semantically meaningful values for each component. It then systematically recombines these values to construct mock responses. By explicitly seeking diversity at the component level, \projName{} uses each generated response to explore a different part of the API's response space.

\projName{} uses these diverse responses to perform \emph{multi-trajectory testing}. During a test, \projName{} intercepts calls to external APIs and supplies generated responses in place of the real service. By repeatedly executing the same test while varying these responses, \projName{} explores how the application behaves under different behaviours of its dependencies. These include common responses, unusual combinations of values, and failures such as timeouts and server errors, providing a form of chaos testing~\cite{chaosmonkey,coyote}. Multi-trajectory testing can therefore exercise application behaviours that a small, fixed set of manually authored mock responses would miss.

Behavioural diversity is important not only for increasing coverage, but also for detecting accidental assumptions about dependencies. Tests built around a small set of fixed mock responses can allow application code to depend on properties of those responses that the API does not guarantee~\cite{mockingstudy}. \projName{} can instead generate \emph{spec-anticipatory} responses: responses that differ from those produced by the current service implementation but remain consistent with its API specification. Testing against such responses can reveal assumptions that hold for the current implementation or mock, but need not hold for other implementations or future versions of the service, where such assumptions can lead to difficult-to-diagnose failures~\cite{semverbugs}.

We implement these ideas in \projName{} and evaluate the resulting system on ten RESTful service benchmarks. 
When compared with state-of-the-art test generation using developer-written mocks or live services, \projName{} 
improves line coverage by up to $12$ percentage points on three benchmarks while achieving comparable coverage on the remaining seven, 
exposes more distinct failures on every benchmark, including an increase from 7 to 22 on one benchmark. Finally, generating mocks with \projName{} 
costs an average of less than $\$0.20$~USD and takes about 2 to 54 minutes across all the benchmarks.\\

\noindent
The key contributions of this work are:

\begin{itemize}
    \item We introduce a novel system (\projName) that automatically generates high-quality mocks for cloud APIs and service-based applications using a 
    combination of LLMs and symbolic techniques. 
    \item We demonstrate how the use of large-language models, symbolic techniques, and the expressive specifications can 
    be leveraged to, reliably, generate a diverse range of mock behaviors (\Cref{sec:llm}).
    \item We present a specialized technique for generating mocks for database-backed applications by translating SQL queries into specifications 
    and leveraging the power of our combined LLM and symbolic techniques (\Cref{sec:dbspec}).
    \item We demonstrate how to create multi-trajectory tests (\Cref{sec:multiworld}) that sample across the space of possible API behaviors for a single test input, to enhance test 
    coverage, ensure failure modes are tested, uncover edge cases, and reduce the risk of introducing accidental assumptions about API behavioral guarantees.
    \item We evaluate on a comprehensive set of RESTful service evaluation benchmarks. On these benchmarks, \projName improves test coverage when compared to 
    traditional hand-authored mocks by up to 8 to 12 percentage points on 3 benchmarks and meets human levels on the remaining cases, and based on the efficiency of the generated mocks, 
    does so at under $\$0.20$~USD per service (\Cref{sec:experiments}).
    \item We leverage the fact that \projName generates a set of fully isolated and serialized tests to enable totally local runs. Our evaluations 
    show (\Cref{sec:experiments}) that we can reduce test execution time (\ie in CI runs) by up to $15\times$ and entirely eliminate the issue of flakey tests.
\end{itemize}

\section{\projName Workflow Overview}
\label{sec:algorithm}
%TeX root = mockmaster.tex

This section provides an overview of the design of the system, reviews prior work on automated \emph{test-input} generation used by \projName, and then introduces our running 
example which we use to illustrate the how the system automatically generates mocks for external data-sources and services.

\subsection{API Specification and Running Example}
A critical issue in framing the problem of automated test generation is the choice which API specification language to use, as it directly impacts the expressiveness and precision of the generated tests 
as well as the practical utility of the resulting system. The most common specification language for RESTful APIs in use today is OpenAPI~\cite{openapi}. However, as systems have grown more complex and 
the quality bar has increased, there is increasing interest in more expressive and precise specification languages that can better capture the nuances of modern APIs -- in industry Microsoft's TypeSpec~\cite{typespec} 
and Smithy~\cite{smithy} from Amazon AWS and in technologies like Lean~\cite{lean} or BAPI~\cite{bapispecs} coming out of academic research. In this work, we use TypeSpec as our API specification language and build our test generation framework and use a simple OpenAPI-To-TypeSpec converter\footnote{
Symbolic OpenAPI-to-TypeSpec converters \url{https://typespec.io/docs/emitters/openapi3/cli/} can be used for direct translation. Alternatively emerging techniques for Agent driven specification extraction 
can be employed to create enriched TypeSpec specifications or to target more expressive systems~\cite{bapispecs,lean}.} for compatibility with existing OpenAPI specifications.

For our running example we consider a conceptual e-commerce application from Morgan Stanley that responds to user requests for renting beach equipment based on expected weather conditions 
and predicted inventory availability. The core of the application logic (implemented in TypeScript) is shown in \Cref{fig:rental_code} and the TypeSpec definitions for the weather forecast API and database queries are shown in \Cref{fig:api-spec-example}.

\begin{figure}[t]
\centering
\begin{lstlisting}[]
scalar Fahrenheit extends int32;
scalar Mph extends int32;

model TempRange { low: Fahrenheit; high: Fahrenheit; }
model WindSpeedRange { min: Mph; max: Mph; }

enum Detail {
    HighWind,
    Thunderstorm,
    HighSurf, //perhaps added in an update
    Sunny,
    Cloudy,
    ...
}

model Forecast {
    temp: TempRange;
    windSpeed: WindSpeedRange;
    stormWatch: boolean;
    ...
    info: Detail[];
}

/** Get the weather forecast for tomorrow for a given city. **/
op getForecast(city: string): Forecast;

/** Get the temperature range for a given city. **/
op getTempRange(city: string): TempRange;

op getInventory(city: string): int32;
op completeReservation(city: string, quantity: int32): string;

\end{lstlisting}
\caption{Weather Forecast and Inventory Control specification (and API) in TypeSpec~\cite{typespec}.}
\label{fig:api-spec-example}
\end{figure}

The code in \Cref{fig:api-spec-example} shows the TypeSpec definitions for the weather forecast API and inventory control API, including the models, enums, and API endpoints used in the rental service implementation.

The code in \Cref{fig:rental_code} shows the implementation of the rental service, which makes use of the weather forecast and inventory control APIs. 
The first action in this code is to make a call to the \cf{getForecast} API (perhaps implemented by wrapping the National Weather Service API~\cite{nationalweatherservice}) to retrieve the weather forecast 
for the specified location. The retrieved forecast is then checked for storm warnings and unsafe wind conditions before proceeding with inventory checks and reservation completion. As a safety 
measure there is also an assertion, just before completing the reservation, ensuring that only sunny or cloudy weather conditions are permitted for rental operations. As with many 
production systems, it is crucial to have such backstop safety checks in place to catch bugs that may be introduced during code development or evolution. 

\begin{figure}[t]
\centering
\begin{lstlisting}[]
function handleRental(city: string, quantity: number): string
{
    await forecast = getForecast(city);
    if(forecast.stormWatch || checkUnsafeWinds(forecast.windSpeed)) {
        return "Request cannot be fulfilled due to unsafe weather conditions.";
    }

    await inventory = getInventory(city);
    if(inventory < quantity) {
        return "Request cannot be fulfilled due to insufficient inventory.";
    }

    // Additional business logic ...

    //Safety is our priority -- no predictions other than sun or clouds!
    assert(forecast.info.every(v => v === Detail.Sunny || v === Detail.Cloudy)); 

    return await completeReservation(city, quantity);
}
\end{lstlisting}
\caption{Rental service implementation which depends on calls to external service APIs (getForecast) and data sources (getInventory which retrieves inventory information from a MongoDB database).}
\label{fig:rental_code}
\end{figure}

In this example there is a bug where the initial forecast safety check is almost correct but, \eg due to an oversight during a refactor, is not comprehensive enough to catch all unsafe weather conditions. 
In fact this code could run in production for an extended period of time before the specific set of conditions needed to trigger it are encountered. A quality test suite could expose this issue prior to deployment, 
but there are several aspects of this sample code that make writing comprehensive tests challenging -- the reliance on external APIs and dynamic data sources, and the need to satisfy a specific set of 
(unusual) conditions that trigger the assertion.

The first challenge is the presence of the external weather forecast API, which introduces external variability and unpredictability in the test environment. The second challenge is 
use of an external data-source which introduces yet another external source of behavior. Finally, there is the actual bug that we want to expose, which occurs on specific combinations of weather conditions 
that may only appear under unusual circumstances -- a day in which the weather service forecast indicates conditions that are storm-free and with low wind conditions but which 
would fail the safety assertion. Perhaps a situation with a day exhibiting high surf conditions from a distant storm but with no local warnings or high winds.

The previous state-of-the-art for testing this code involves either 1) running against live services and accepting the variable (flakey~\cite{flakeytestexperience}) nature of the resulting test suite, 2) manually 
mocking the external services/data sources to create controlled test scenarios and paying the engineering costs associated with maintaining and updating these mocks~\cite{mockingstudy}, or 3) ``testing in production'' 
and planning to have enough telemetry in place to detect and respond to issues quickly. All of these existing approaches are unattractive, and as we demonstrate there is actually an option $\#4$, which is to automatically 
generate a high-quality set of diverse mock values that comprehensively and deterministically test the and-to-end behavior the application.

\subsection{Test-Input Generation}
For the purposes of \projName, we can use any form of test-input generation technique to create initial test cases that will later be used to drive the automated mock generation process -- this can range from 
existing human authored tests, to classic fuzzing techniques~\cite{afl}, to genetic or reinforcement learning based input generators~\cite{tecton,arat-rl,evomaster}. 

\begin{table}
\caption{Example input test cases for the \cf{handleRental} API.}
\label{tab:handleRental-inputs}
\centering
\ttfamily
\begin{tabular}{cc}
\toprule
\textbf{city} & \textbf{quantity} \\
\midrule
Long Beach & 1 \\
Huntington Beach & 2 \\
Huntington Beach & 0 \\
\ldots & \ldots \\
Newport Beach & 4 \\
\bottomrule
\end{tabular}
\end{table}

In our running example, \Cref{fig:api-spec-example,fig:rental_code}, the 
input structure for \cf{handleRental} is a simple location (as a string) and number, both of which can be easily generated in a comprehensive manner by the test-input generation techniques mentioned above.
A partial set of possible set of inputs for the \cf{handleRental} function is shown in \Cref{tab:handleRental-inputs}.
These input test cases illustrate a variety of possible scenarios for the \cf{handleRental} function, including typical rentals and edge cases such as zero quantity.

\begin{figure}
    \centering
    \resizebox{.7\linewidth}{!}{%
    \begin{tikzpicture}[
        font=\small,
        >={Stealth[length=2mm]},
        application/.style={draw=orange!65!black, fill=orange!4, rounded corners=3pt, line width=0.8pt},
        cache/.style={draw=blue!60!black, fill=blue!3, rounded corners=3pt, line width=0.8pt, inner sep=9pt, align=left},
        call site/.style={draw=orange!45!black, fill=white, rounded corners=2pt, minimum width=3.8cm, minimum height=0.65cm, font=\footnotesize\ttfamily},
        call/.style={->, draw=orange!75!black, line width=1pt},
        execution/.style={->, draw=green!45!black, line width=1pt},
        edge label/.style={font=\scriptsize, fill=white, inner sep=2pt, align=center}
    ]
        \path[application] (0,0) rectangle (4.4,6.2);
        \node[font=\bfseries] at (2.2,5.8) {Rental application};
        \node[align=center] (inputs) at (2.2,7.15) {
            Test inputs: $t_1,\ldots,t_n,\ldots$\\[2pt]
            \textbf{Selected:} \texttt{(HB, 2)}
        };
        \draw[execution] (inputs.south) -- (2.2,6.2);
        \node[font=\footnotesize] at (2.2,5.3) {\texttt{city = HB}, \texttt{quantity = 2}};

        \node[call site] (forecast) at (2.2,4.7) {getForecast(city)};
        \node[call site] (inventory) at (2.2,2.35) {getInventory(city)};
        \draw[execution] (2.2,5.1) -- (forecast.north);

        \node[align=center] at (9.35,5.9) {
            \textbf{\projName}\\[2pt]
            Intercepted calls and mock-response caches
        };
        \node[cache, minimum width=5.8cm, minimum height=2.1cm] (forecast cache) at (9.35,4.2) {
            \begin{tabular}{@{}l@{}}
                \textbf{Forecast candidates for HB}\\[3pt]
                \{\ldots, no-storm, ["Sunny"] \}\\
                \{\ldots, storm-warning, ["Thunderstorm"] \}\\
                Timeout\\
                \ldots
            \end{tabular}
        };
        \node[cache, minimum width=5.8cm, minimum height=1.75cm] (inventory cache) at (9.35,1.85) {
            \begin{tabular}{@{}l@{}}
                \textbf{Inventory candidates for HB}\\[3pt]
                $0,\quad 3,\quad 10,\quad \ldots$
            \end{tabular}
        };

        \draw[call] (forecast.east) -- (forecast cache.west |- forecast.east)
            node[midway, above=3pt, edge label] {\texttt{getForecast(HB)}};
        \draw[execution] ($(forecast cache.west)+(0,-0.55)$) -- (2.2,3.65)
            node[pos=0.3, below=3pt, edge label] {Selected: \texttt{no-storm \& Sunny}}
            -- (inventory.north);
        \node[font=\footnotesize, fill=orange!4, inner sep=3pt] at (2.2,3.05) {Weather checks pass};

        \draw[call] (inventory.east) -- (inventory cache.west |- inventory.east)
            node[midway, above=3pt, edge label] {\texttt{getInventory(HB)}};
        \draw[execution] ($(inventory cache.west)+(0,-0.6)$) -- (2.2,1.25)
            node[pos=0.3, below=3pt, edge label] {Selected: \texttt{10}}
            -- (2.2,0.65);
        \node[font=\footnotesize] at (2.2,0.35) {Further execution \ldots};
        \node[align=center] (result) at (2.2,-0.8) {
            Test result\\[2pt]
            \texttt{"Confirmed"}
        };
        \draw[execution] (2.2,0) -- (result.north);
        \node[align=center, font=\footnotesize] at (9.35,0.25) {
            On cache misses \Cref{alg:gen-tests} is used\\ 
            to generate candidate responses
        };
    \end{tikzpicture}%
    }
    \caption{Test execution with automatically generated mocks. The selected input uses Huntington Beach (HB) and quantity 2. \projName intercepts external API calls and returns candidate responses from per-API caches, generating candidates on cache misses. Orange arrows show API calls; green arrows show execution and selected mock responses.}
    \label{fig:architecture}
\end{figure}
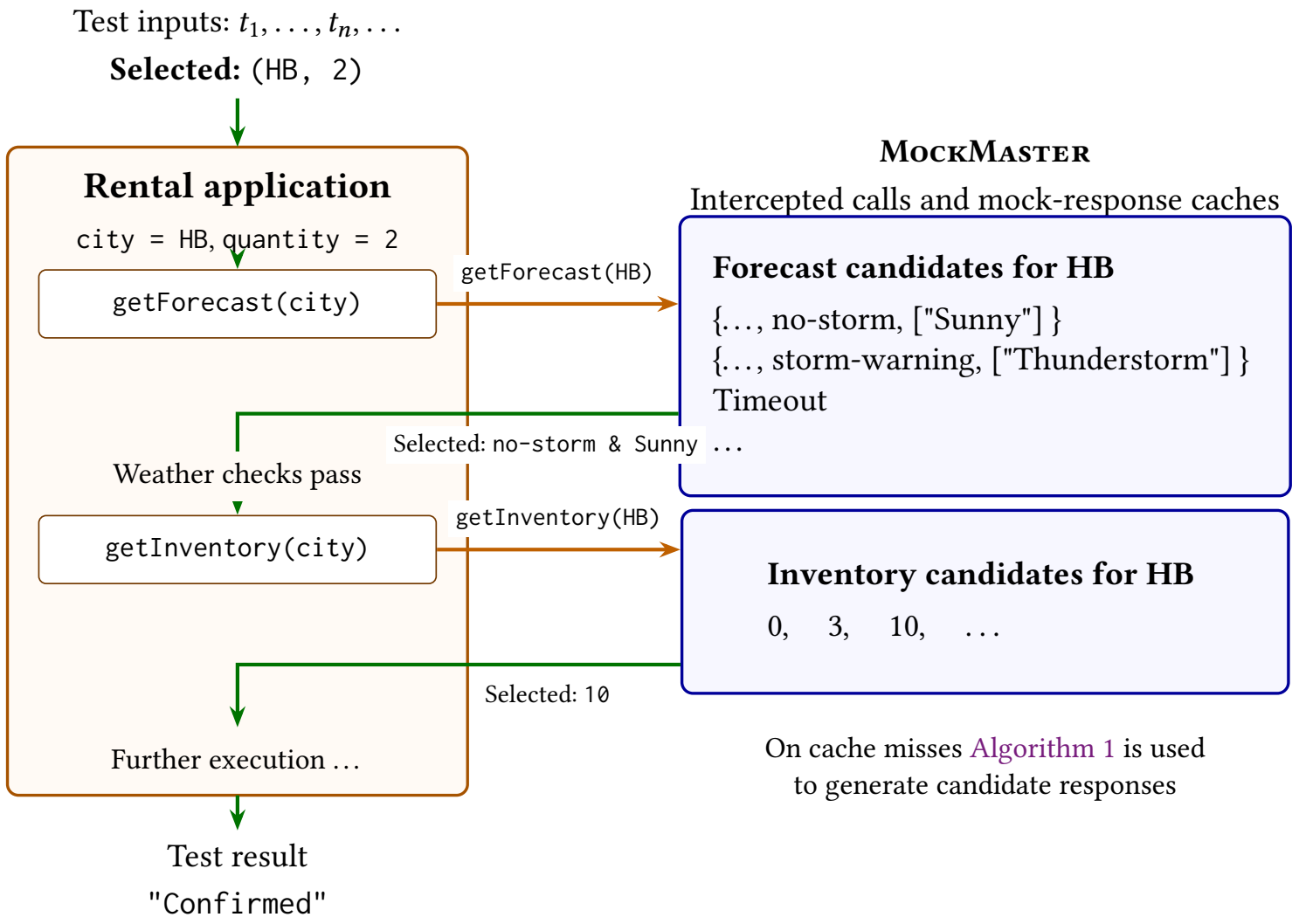

\subsection{Automated Mock Generation with LLM-Executed APIs}
From this set of initial test inputs, a developer would normally need to write mock data manually for each of the external APIs that the application interacts with, ensuring that for each input test case and 
for any values it calls the API with, there is a data entry that corresponds to the expected response from the API. As this task is time consuming and error-prone, human developers often end up producing 
small mock datasets where the entries are \emph{happy path results}\footnote{Happy-path tests omit edge and failure cases -- instead only covering common expected and successful interactions.} and often 
designed so that a single entry can satisfy multiple test cases, reducing the overall effort required to maintain the mock dataset but with the secondary effect of reducing the effectiveness of the test scenarios.

The \projName algorithm operates by hooking into each API call in the application. As shown in \Cref{fig:architecture,tab:getForecast-cache}, whenever a test is run and a external API (or resource) 
is needed the call, and concrete argument values, are captured and \projName uses the algorithm in \Cref{alg:gen-tests} to generate a range of candidate mock responses which encompass a range of possible outcomes for the given arguments.

\begin{table}
\caption{Example cache of generated mock values for the \cf{getForecast} API.}
\label{tab:getForecast-cache}
\centering
\ttfamily
\begin{tabular}{ll}
\toprule
\textbf{City} & \textbf{Forecast} \\
\midrule
Huntington Beach & \{ \{56, 72\}, \{0, 12\}, false, ["Sunny"] \} \\
Huntington Beach & \{ \{61, 78\}, \{5, 10\}, true, ["Cloudy", "Thunderstorm"] \} \\
\ldots & \ldots \\
Huntington Beach & \{ \{71, 85\}, \{5, 15\}, false, ["Sunny", "HighSurf"] \} \\
Huntington Beach & \cf{HTTP 500} \\
\ldots & \ldots \\
Newport Beach & \{ \{64, 75\}, \{20, 40\}, true, ["HighWind"] \} \\
Newport Beach & \{ \{45, 64\}, \{5, 7\}, false, [] \} \\
Newport Beach & \cf{TIMEOUT} \\
\ldots & \ldots \\
\bottomrule
\end{tabular}
\end{table}

\Cref{fig:architecture} shows the overall architecture and flow the \projName system when run on a test-input. In this illustration the system is in the process of executing the test corresponding to 
with the input values \cf{city="Huntington Beach"} (HB) and \cf{quantity=2}. As the code executes and reaches the first API call, to \cf{getForecast}, instead of executing the call to the actual API,
\projName intercepts the call and generates a mock response based on the captured arguments and the candidate mock responses. 

As shown in \Cref{tab:getForecast-cache}, the system maintains a cache of generated mock responses to improve efficiency and provide a variety of possible outcomes for an API call. In our example case this cache contains 
(or we can generate) several mock responses for the "Huntington Beach" city input -- these include cases where the weather is sunny and storm free (a happy path), cases where there is a \cf{stormWatch}, as well 
as the unusual case (which triggers our assert) of a warm sunny day but with dangerous surf. The \projName system can also generate failure (chaos~\cite{chaosmonkey,coyote}) responses such as timeouts, network 
failures, or error responses such as HTTP 500.

The \projName system will first check if there are any cached mock responses for the given API call. If there are then it will simply select (via controlled random exploration) from the possible matches. If not then it uses the algorithm 
shown in \Cref{sec:llm,sec:dbspec} to populate the cache with new values. It will then construct the return value and provide it to the calling code which will continue execution as if the 
API call had been made normally. This process is then repeated for each subsequent API call encountered during the test execution. As each API may have many possible responses we can re-run a given input multiple 
times, with different mock responses used each time, to do multi-trajectory testing and enhance coverage of possible API behaviors (\Cref{sec:multiworld}). We repeat this process for each test input to systematically 
explore a comprehensive set of possible program executions.

In our running example the system might select the happy-path value \cf{\{ \{56, 72\}, \{0, 12\}, false, ["Sunny"] \}} for the result of the \cf{getForecast} API call and, as a result, we would cover the code 
associated with a successful rental and confirmation of the booking. On a subsequent run, the system might select a different mock response, \eg the anomalous \cf{\{ \{71, 85\}, \{5, 15\}, false, ["Sunny", "HighSurf"] \}} 
result, which would catch the assertion failure. And even later runs might select the 500 response -- ensuring testing of the application's handling of server errors.

\section{API Response Generation Algorithm}
\label{sec:llm}
%TeX root = mockmaster.tex

This section details the core algorithm used by \projName to generate mocked responses for arbitrary RESTful API calls. As database operations are a common and special case 
API, \Cref{sec:dbspec}, describes how \projName specializes its response generation for database interactions when we have additional knowledge about database schema and 
query semantics.

Our objective is, when given the arguments for a call to an API, to generate a plausible response using \emph{only} the information available in the API specification and the input arguments. 
A natural first attempt for solving this problem might be use directly prompt an LLM to (one-shot) generate the response object~\cite{toolemu,toolalpaca} or to use a coding agent to implement 
a code-stub mock implementation that can be executed to produce the response~\cite{ning2026executable}. These techniques are highly effective when the API argument/result values are small 
(low context) and the API behavior is relatively simple. However, similar to issues seen when directly prompting LLMs to generate complex test \emph{API inputs}~\cite{ning2026executable,teststruggle}, 
when the API outputs become larger and have more intricate response structures, these approaches become less effective and suffer from limited variety in the 
output generated.

To counteract this tendency we adopt a similar approach to \emph{output} generation as is used by \tecton for ensuring diverse coverage during \emph{input} generation. The algorithm 
shown in \Cref{alg:gen-tests} illustrates this approach. Instead of using a purely AI driven approach, this algorithm combines a symbolic decomposition of the response type into its 
constituent components (\cf{getComponents}), and then uses AI-driven value generation for each component (\cf{generateValues}), a (neuro)symbolic recombination of these component values 
into a complete candiate response object (\cf{genSuccessOpts}), and finally to ensure coverage of failure scenarios (\cf{genFailureOpts}) is used to generate a set of possible failure responses.

\begin{algorithm}[t]
\caption{Generating a set of possible API, success and failure, response values with \projName.}
\label{alg:gen-tests}
\begin{algorithmic}
\Function{ResultGen}{\emph{args}: String[], \emph{t}: Type} \Comment{Returns output option success/failure tuple}
    \State components $\gets$ \Call{getComponents}{t, 'res'}
    \ForAll{c $\in$ components}
        \State c.values $\gets$ \Call{generateValues}{\emph{args}, c}
    \EndFor

    \State rsuccess $\gets$ \Call{genSuccessOpts}{\emph{args}, components}
    \State rfailure $\gets$ \Call{genFailureOpts}{\emph{args}}

    \State \Return (rsuccess, rfailure)
\EndFunction
\end{algorithmic}
\end{algorithm}

The \cf{getComponents} in \Cref{alg:gen-tests} recursively traverses a type structure to identify all primitive components:  properties, array entries, or scalar declarations of primitive 
types (numbers, strings, booleans, etc.). These primitives form the atomic output spaces for subsequent test generation. \textsc{GetComponents} tracks a \emph{path}: a dot-separated string 
(\eg, "res.temp.low.value" from the running example) that records the location of each primitive from the root of the structure. Using this recursive decomposition and path tracking 
enables the \cf{getComponents} function to build up a (named) list of all the primitive values, their types, and corresponding paths needed to construct a complete response object. In 
the case of the \cf{TempRange} type, this results in components shown shown in \Cref{tab:temp-range-components} -- which has 2 primitive components corresponding to the low and high values 
of the \cf{TempRange} model, each of which is of type \cf{int32}, and at this stage, no values have been generated for them yet.

\begin{table}[t]
\caption{Computed components for the \cf{TempRange} type -- prior to generating representative values.}
\label{tab:temp-range-components}
\centering
\ttfamily
\begin{tabular}{lcc}
\toprule
\textbf{Path} & \textbf{Type} & \textbf{Values} \\
\midrule
\cf{v.low.value} & \cf{int32} & $\emptyset$ \\
\cf{v.high.value} & \cf{int32} & $\emptyset$ \\
\bottomrule
\end{tabular}
\end{table}

The \cf{generateValues} function in \Cref{alg:gen-tests} takes the computed components and produces representative values for each primitive component to fill out their \cf{values}. 
\projName uses a simple prompting structure for LLM generation of these values, based on classic black-box (or combinatorial) test construction~\cite{combtesting}, 
that provides a general prompt, the specifications of all relevant types, the input arguments for the API call, and then the path/type of the specific primitive for which values are being generated. A sample of this 
prompt\footnote{This prompt has been simplified for space and clarity. The complete prompting scheme is included in the reproduction materials.} is shown, in \Cref{fig:sample-prompt} 
for generating the values for the  \cf{v.low.value} component.

\begin{figure}[t]
    \footnotesize
    \centering
    \begin{tcolorbox}[
        colback=white, % Background color of the box
        colframe=black, % Outline color of the box
        boxrule=1pt, % Thickness of the box outline
        arc=4mm, % Radius of the rounded corners
        title=LLM Prompt Context for \cf{v.low.value} when the argument is \cf{"Huntington Beach"}., % Title for the box
        fonttitle=\bfseries % Make the title bold
    ]
    System:\\
    You are an expert in generating test values for software components. We are generating a mock stub for an API. 
    You are given the API, a set of type definitions, and a path to a specific primitive component in the output 
    value, and an input that the API is invoked with.\\
    
    Your job is to generate a JSON array containing ONLY values strictly matching the specified data type and format 
    that could be possible values for the given component in the output. The values in the array should be diverse, 
    covering all possible interesting value equivalence classes as well as boundary conditions.\\

    Context:\\
    API: \cf{api getTempRange(city: string): TempRange;}\\
    DocComment: \cf{"/** Get the temperature range for a given city. **/"}\\
    Types: \cf{scalar Fahrenheit extends int32; model TempRange \{ low: Fahrenheit; high: Fahrenheit; \}}\\
    Path: \cf{res.temp.low;}\\
    Args: \cf{"Huntington Beach"}\\
    \end{tcolorbox}
    \caption{Sample LLM prompt for generating values for the \cf{v.low.value} component.}
    \label{fig:sample-prompt}
\end{figure}

A key aspect of this prompt is that it is focused on the task of generating value for a \emph{single} component of the 
result \emph{not} the \emph{entire} output. As seen in the prior \tecton~\cite{tecton} work this approach results 
in much more effective coverage of interesting test values and allows for much lower LLM inference costs than directly 
generating entire outputs at a time. In addition the explicit information on what inputs this output corresponds, and 
the information on the intent and expected behavior of the API ensures that the generated values are meaningful and 
relevant to the actual use case.

\begin{table}[t]
\caption{LLM generated values for \cf{v.low.value} and \cf{v.high.value} in \cf{TempRange}.}
\label{tab:temp-range-llm-example}
\centering
\ttfamily
\begin{tabular}{lcc}
\toprule
\textbf{Path} & \textbf{Type} & \textbf{Values} \\
\midrule
\cf{v.low.value} & \cf{int32} & $\lbrace 21, 32, 60, 82 \rbrace$ \\
\cf{v.high.value} & \cf{int32} & $\lbrace 40, 70, 84, 95, 110 \rbrace$ \\
\bottomrule
\end{tabular}
\end{table}

\Cref{tab:temp-range-llm-example} shows generated values for the \cf{v.low.value} and \cf{v.high.value} components in 
the \cf{TempRange}. Note that the generated values cover a range of plausible temperatures and are representative of 
both expected (and boundary condition) values that one would expect for weather in a southern California beach city 
like Huntington Beach.

The final step in the process of generating test values for the \cf{TempRange} model is to combine these LLM-generated 
primitive values from each component to create the full composite values needed for the mock implementation. There are 
multiple possibilities for this operation, including random, combinatorial generation, LLM driven, \etc, and we leave 
a comprehensive evaluation and comparison of these to future work. Instead we take a simple, and effective, baseline 
approach of constructing one composite value at a time until we have used every generated primitive value at least once.
This ensures that all generated primitive values are covered in the final set mock value choices while, as per the evaluation (\Cref{sec:experiments}), 
producing a set of mock results that is both comprehensive in terms of covering the generated primitive values and 
manageable in size.

\begin{table}[t]
%\begin{wraptable}{r}{0.45\linewidth}
\caption{Example mock result set for the \cf{TempRange} model by value saturation using the values from \Cref{tab:temp-range-llm-example}
}
\label{tab:temp-range-one-way}
\centering
\ttfamily
\begin{tabular}{ccc}
\toprule
\textbf{v.low.value} & \textbf{v.high.value} & \textbf{TempRange} \\
\midrule
21 & 40 & \{low: 21, high: 40\} \\
32 & 40 & \{low: 32, high: 40\} \\
60 & 95 & \{low: 60, high: 95\} \\
82 & 70 & \{low: 82, high: 70\} \\
60 & 84 & \{low: 60, high: 84\} \\
82 & 110 & \{low: 82, high: 110\} \\
\bottomrule
\end{tabular}
\end{table}
%\end{wraptable}

A possible set of mock results is shown in \Cref{tab:temp-range-one-way}. In this table we see that each value 
for \cf{low} and \cf{high} has been used at least once in the mock result set. Further, the combinations include 
a range of possible quite plausible (low of 60 high of 84) and extreme (low of 21 high of 40) combinations. It 
is also possible that some combinations may not be semantically meaningful, such as the low temperature (82) that 
is higher than the high temperature (70). We do not attempt to filter these out here but instead rely on the API 
data-validation logic to catch and handle such cases if they are encountered\footnote{Developments in improved 
API specification languages is an open research area and developments like executable pre/post conditions on 
APIs~\cite{bapispecs} or dependent logical types~\cite{lean} are a promising direction for addressing such issues.}.

\section{Database API Specialization}
\label{sec:dbspec}
Given the prevalence of database operations in modern applications and the extensive experience LLMs have 
with understanding DB schema, queries, and database operations, \projName provides a custom path for 
mock generation when working with them. 

If a database is used as an external dependency, \projName attempts to locate a schema declaration for 
the database. In the case of SQL a DDL or for NoSQL flavors we look for POJO (Plain Old Java Objects are 
often used in Spring Applications) or JSON Schema specifications. Then, if \projName is able to locate the schema, 
it will leverage the schema and the exact query syntax to augment the standard value generation prompt as 
described in \Cref{sec:llm} to generate contextually accurate mock data.

Thus, if in the running example the \cf{inventory} Mongo database is defined with a POJO schema, then that 
schema, and the actual query text from the application are passed to the LLM as part of the prompt:
\begin{lstlisting}[language=Java]
public class Inventory {
    public String city;
    public int quantity;
    public double rate;
}

eq("city", "Huntington Beach")
\end{lstlisting}

With this additional context the LLM is able to generate semantically accurate and 
contextually relevant mock data for the query results. In our running example the 
mock result for the query might produce the following document result options:
\begin{lstlisting}
[
    {
        "city": "Huntington Beach",
        "quantity": 10,
        "rate": 29.99
    },
    {
        "city": "Huntington Beach",
        "quantity": 0,
        "rate": 19.99
    }
    ...
]
\end{lstlisting}

Selecting from these various possible outcomes, as described in \Cref{sec:multiworld}, 
allows \projName to explore a range of these possible outcomes -- in this case the 
scenario where there is ample inventory for the queried city, as well as the scenario 
where the inventory is unavailable for the requested day.

\section{Multi-Trajectory Coverage}
\label{sec:multiworld}
A key feature of the algorithms in \Cref{sec:llm,sec:dbspec} is that they produce multiple possible versions 
of API responses or Database query results. This capability allows us to extend \projName to perform 
multi-trajectory coverage for each test input. In this form the system explores multiple possible execution 
paths that can result from different result values -- normal values \vs extremes or failures \vs successes.

\subsection{Failure Response Generation}
\label{sec:llmfail}
The first step is to augment the LLM generated results in \Cref{sec:llm,sec:dbspec}, associated with successful 
returns, with additional failure variations to cover scenarios such as timeouts, authorization errors, network failures, 
and other exceptional conditions. Testing failure scenarios is notoriously difficult in traditional testing 
approaches and chaos testing these scenarios has been shown to reveals large numbers of (serious) 
issues~\cite{mockingstudy,chaosmonkey,chaosmusuvathi} around untested (or under-tested) error scenarios.

RESTful APIs, and their specifications, often use implicit representations for failure responses, 
including HTTP status codes or special sentinel codes that are placed directly in the response body. 
Thus, instead of trying to roll this generation into the main response generation 
process (\Cref{sec:llm,sec:dbspec}), where the additional request to generate failure values can distract 
the model, \projName generates failure responses as a separate step, ensuring that the exploration of 
error scenarios is thorough and does not interfere with the generation of normal responses.

Given the more limited scope of the error generation task, and the implicit nature of many failure representations, 
we lean more heavily on the capabilities of the LLM for generation, and directly ask the LLM to generate a list of 
possible error responses for the given API call. The LLM is expected to reason about the API specification, the 
types, and the input arguments to enumerate all possible error responses.

\begin{figure}[t]
    \footnotesize
    \centering
    \begin{tcolorbox}[
        colback=white, % Background color of the box
        colframe=black, % Outline color of the box
        boxrule=1pt, % Thickness of the box outline
        arc=4mm, % Radius of the rounded corners
        title=LLM Prompt Context for generating error responses when the argument is \cf{"Huntington Beach"}., % Title for the box
        fonttitle=\bfseries % Make the title bold
    ]
    System:\\
    You are an expert in generating test values for software components. We are generating a mock stub for an API. 
    You are given the API, a set of type definitions, and an input that the API is invoked with.\\
    
    Your job is to generate a JSON array containing ONLY possible ERROR responses for the call. The array should 
    include relevant error scenarios, such as invalid inputs, failed authentication, network errors, or other plausible 
    conditions that could lead to a failure response. \\

    Context:\\
    API: \cf{api getTempRange(city: string): TempRange;}\\
    DocComment: \cf{"/** Get the temperature range for a given city. **/"}\\
    Types: \cf{scalar Fahrenheit extends int32; model TempRange \{ low: Fahrenheit; high: Fahrenheit; \}}\\
    Args: \cf{"Huntington Beach"}\\
    \end{tcolorbox}
    \caption{Sample LLM prompt for generating error responses for a call to \cf{getTempRange} when the argument is \cf{"Huntington Beach"}.}
    \label{fig:sample-error-prompt}
\end{figure}

\projName uses a simple prompting structure for LLM generation that includes a general system prompt requesting only 
error responses, the specifications of all relevant types, and the input arguments for the API call. A (simplified) sample of this 
prompt is shown, in \Cref{fig:sample-error-prompt} for generating the error responses for the API call. 

\begin{table}[t]
\caption{LLM generated error responses for a call to \cf{getTempRange} when the argument is \cf{"Huntington Beach"}.}
\label{tab:temp-range-llm-error-example}
\centering
\ttfamily
\begin{tabular}{lc}
\toprule
\textbf{Code} & \textbf{Value} \\
\midrule
400 & \text{"Invalid city name"} \\
504 & \text{"Timeout"} \\
401 & \text{"Authentication failed"} \\
500 & \text{"Error"} \\
\bottomrule
\end{tabular}
\end{table}

\Cref{tab:temp-range-llm-error-example} shows generated values for the error responses for a call to \cf{getTempRange} 
when the argument is \cf{"Huntington Beach"}. As can be seen, the LLM was aware of possible error types, and how they 
should be communicated as HTTP status codes, and is also able to generate appropriate error messages for each scenario 
based on any DocComments provided for the API.

\begin{figure}
    \centering
    \resizebox{.85\linewidth}{!}{%
    \begin{tikzpicture}[
        font=\small,
        >={Stealth[length=2mm]},
        call site/.style={draw=orange!45!black, fill=white, rounded corners=2pt, line width=0.8pt, minimum width=3.1cm, minimum height=0.65cm, font=\footnotesize\ttfamily},
        result/.style={draw=orange!65!black, fill=orange!4, rounded corners=3pt, line width=0.8pt, minimum width=3.1cm, inner sep=6pt, align=center, font=\footnotesize},
        failure/.style={draw=red!60!black, fill=red!3, rounded corners=3pt, line width=0.8pt, minimum width=3.1cm, inner sep=6pt, align=center, font=\footnotesize},
        execution/.style={->, draw=green!45!black, line width=1pt},
        error/.style={->, draw=red!60!black, line width=1pt},
        edge label/.style={font=\scriptsize, fill=white, inner sep=2pt, align=center}
    ]
        \node[align=center] (input) at (0,1.4) {
            \textbf{Same initial input}\\[2pt]
            \texttt{("Huntington Beach", 2)}
        };
        \node[call site] (forecast) at (0,0) {getForecast(city)};
        \draw[execution] (input.south) -- (forecast.north);

        \node[failure] (network error) at (8.6,0) {Terminated\\503 Network error};
        \draw[error] (forecast.east) -- (network error.west)
            node[midway, above=3pt, edge label] {Failure};

        \node[call site] (inventory failure) at (3.8,-2.2) {getInventory(city)};
        \node[call site] (inventory ten) at (3.8,-3.8) {getInventory(city)};
        \node[call site] (inventory zero) at (3.8,-5.4) {getInventory(city)};
        \draw[draw=green!45!black, line width=1pt] (forecast.south) -- (0,-2.2)
            node[midway, right=3pt, edge label] {Selected:\\\texttt{no-storm \& Sunny}}
            -- (0,-5.4);
        \draw[execution] (0,-2.2) -- (inventory failure.west);
        \draw[execution] (0,-3.8) -- (inventory ten.west);
        \draw[execution] (0,-5.4) -- (inventory zero.west);

        \node[failure] (database error) at (8.6,-2.2) {Terminated\\503 Database error};
        \node[result] (confirmed) at (8.6,-3.8) {Success\\\texttt{"Confirmed"}};
        \node[result] (insufficient) at (8.6,-5.4) {Declined\\\texttt{"Insufficient inventory"}};
        \draw[error] (inventory failure.east) -- (database error.west)
            node[midway, above=3pt, edge label] {Failure};
        \draw[execution] (inventory ten.east) -- (confirmed.west)
            node[midway, above=3pt, edge label] {Selected: \texttt{10}};
        \draw[execution] (inventory zero.east) -- (insufficient.west)
            node[midway, above=3pt, edge label] {Selected: \texttt{0}};
    \end{tikzpicture}%
    }
    \caption{Multi-trajectory exploration tree for the example, illustrating the ability to test failure logic and multiple failure scenarios 
    over multiple trajectories starting from the same initial state.}
    \label{fig:multi-trajectory}
\end{figure}
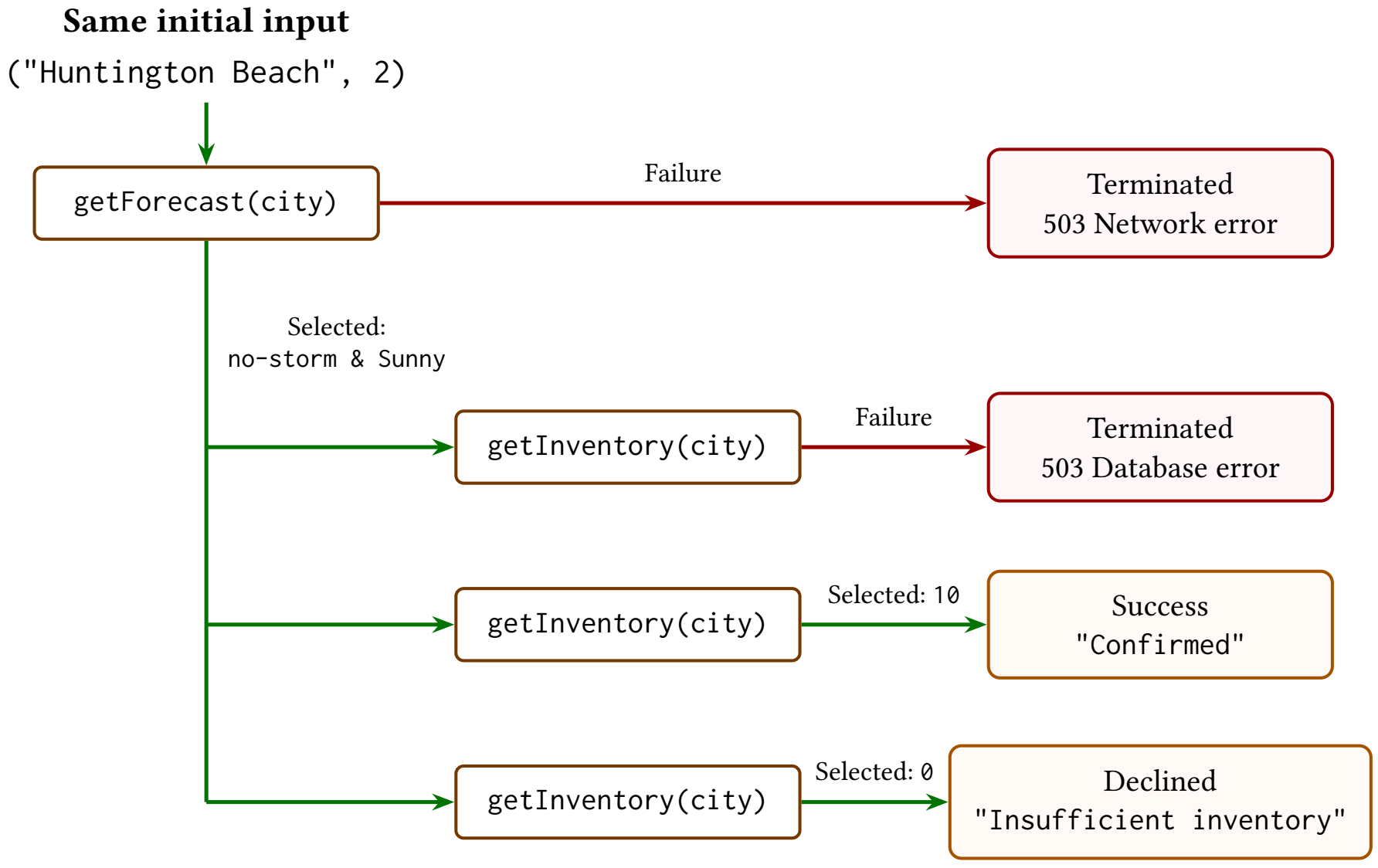

\subsection{Multi-Trajectory Exploration}
Using the LLM value and LLM error generation techniques, along with the caching mechanisms described in \Cref{fig:architecture}, 
we can implement a multi-trajectory exploration framework for API responses. 

Each element in the cache is associated with a specific API call and its corresponding arguments. The value in the 
cache entry is a tuple of lists \cf{(Success[], Error[])}. When an API call is made, the \projName framework intercepts it 
and checks if a matching set of argument values are in the cache. If not then \projName generates a set of possible successful continuations
using the LLM value generation techniques from \Cref{sec:llm,sec:dbspec}, and also, generates a set of 
possible failure continuations based on the technique described in \Cref{sec:llmfail}. These are then stored in the 
cache for efficient reuse.

Once \projName has ensured that the cache contains the needed value (either already existed or was populated) it can 
simply retrieve the corresponding lists. Using these lists \projName makes two randomized exploration decisions based on a
configurable ($p_{\text{failure}}$) parameter and randomized selection within the lists. The first is a decision on whether to 
follow a successful path or to explore an error path. The second decision is the selection of a specific continuation 
from the chosen path's list of possibilities.

For example if $p_{\text{failure}}$ is set to 0.1, then there is a 10\% chance that \projName will choose to explore 
an error path, and a 90\% chance it will follow a successful path. In our example if the random value generated is say, 
$0.5$, the \projName will opt to follow the success path. It will then look at the list of possible successful 
continuations. 

Here it will use a randomized choice of the specific continuation to follow. In our example it may select a 
result with a Sunny day and no storms, say index $0$ in \Cref{tab:getForecast-cache}, and will marshal the value into the appropriate format for the 
application to consume and then continue execution. The current implementation of \projName uses a simple 
fully randomized approach for this selection. In many applications this is sufficient, but more sophisticated 
least-recently-used (LRU) or using history or prior API responses to guide the selection can be employed to 
improve exploration.

To drive the actual exploration, \projName runs each test input 3 times with a fixed set of $p_{\text{failure}}$ 
values ranging over [0.10, 0.30, 0.50]. Based on our experimental results, this design provides 
effective coverage and exploration, while the purely randomized approach allows for zero-coordination parallel 
execution of tests across multiple instances or threads.

\Cref{fig:multi-trajectory} illustrates the multi-trajectory exploration tree for the running example, highlighting 
how \projName can explore tests, both successful and failure paths, across multiple trajectories. This figure shows 
multiple trajectories all starting with the same initial input, \cf{"Huntington Beach"} and then branching out into 
different trajectories based on various mock continuations.

In the first case \projName explores the path where the call to getForecast immediately fails with a network error 
and the application terminates. The next trajectory explores the path where the call to getForecast succeeds but the 
attempt to read the inventory from the database fails, leading to a different failure trajectory. The next two 
trajectories explore scenarios where the mocked calls on the paths succeed but the result values differ, thus 
leading to different end states for the application, and allowing \projName to thoroughly explore the space of 
possible application states and behaviors.

\section{Evaluation}
\label{sec:experiments}

This section presents the results of the empirical studies conducted to assess \projName. Our evaluation aims to
address the following research questions:
\begin{description}
\item[RQ1:] How do the \projName results compare with application-provided mocks and live-service backed executions \wrt code coverage and error detection?
\item[RQ2:] What is the cost, in both time and LLM inference spend, of using \projName for test generation?
\item[RQ3:] What is the effect of key \projName features on detection of edge-case or intermittent failure bugs compared to application-provided mocks?
\item[RQ4:] As \projName enables the replacement of live services with mocks, how does this impact the overall testing experience in terms of flakiness and test-suite execution time?
\end{description}

\subsection{Methodology}

\begin{table}
\caption{RESTful service benchmarks statistics --- Source LOC, total number of API endpoints, and number of mocked dependencies (Database and REST) in each service.}
\label{tab:benchmarks}
\centering
\begin{tabular}{lrrrr}
\toprule
REST Service      & Code Lines & Endpoints & Mocked DB & Mocked REST \\
\midrule
CWA      & 3.9k      & 5 & 1 & 1 \\
Cat Watch            & 9.6k       & 6 & 1 & 1 \\
Feature Service  & 1.6k       & 18 & 1 & 0 \\
Genome Service & 22.1k       & 46 & 1 & 5 \\
PetClinic & 10.5k     & 34 & 1 & 0 \\
User Management   & 2.8k       & 14 & 1 & 0 \\
Project Tracking   & 3.7k      & 67 & 1 & 0 \\
Person Controller & 0.6k       & 12 & 1 & 0 \\
Market & 7.9k       & 13 & 1 & 0 \\
Rental Service & 2.9k       & 2 & 1 & 1 \\
\bottomrule
\end{tabular}
\end{table}

\begin{figure}
    \centering
    \begin{minipage}{\linewidth}
        \centering
        \includegraphics[width=\linewidth]{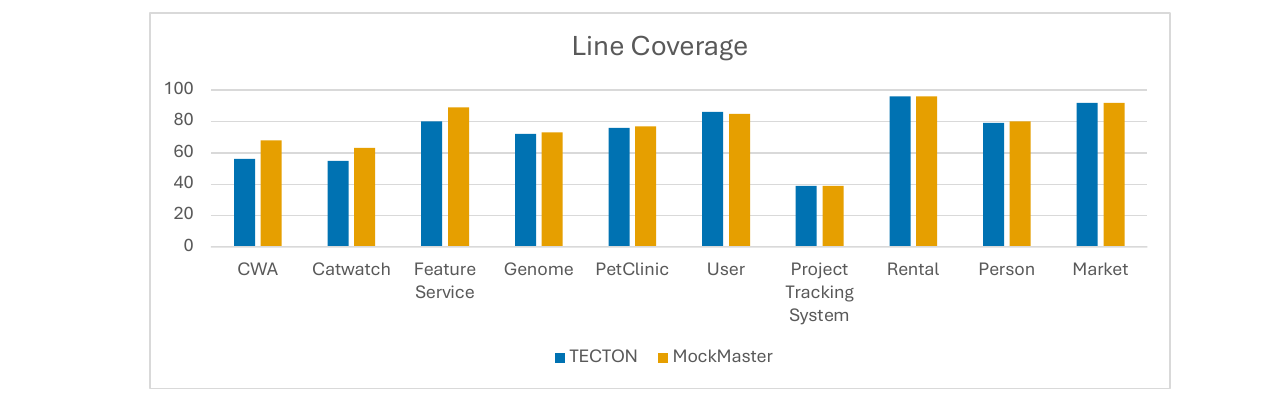}
    \end{minipage}\\
    \begin{minipage}{\linewidth}
        \centering
        \includegraphics[width=\linewidth]{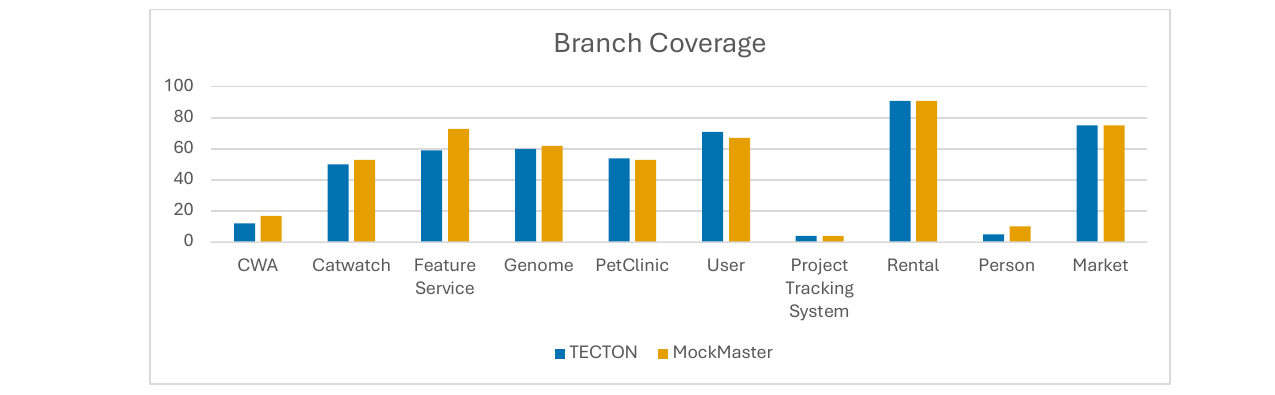}
    \end{minipage}
    \caption{Comparison of \tecton with application-provided mocks/live services (baseline) \vs \projName on line and branch coverage.}
    \label{fig:models_comparison}
\end{figure}

This evaluation uses the publicly available RESTful benchmarks introduced by Decrop \etal~\cite{prab}, which 
includes benchmarks that have been used to evaluate numerous RESTful API testing tools~\cite{autoresttest,arat-rl,emtest,llamaresttest,evomaster}.
As our focus is generating mocks for RESTful API testing, we selected $\totalbenchmarks$ Java Spring Boot applications from these benchmarks based on four criteria: The selected application must 1) use at least one external service or data source, 2) have publicly available source code, 3) be  
usable, including all its dependencies, without payment, and 4) utilize 2 or more REST verbs (GET, POST, PUT, and DELETE). Our data set includes 
the \cf{Rental Service} based on the Morgan Stanley tech demo application from the running example (\Cref{fig:rental_code}). 
It also includes 4 benchmarks, \cf{Person Controller}, \cf{Market Rest}, \cf{User Management}, and  \cf{Project Tracking}, which also meet our criteria and have been heavily used in prior work~\cite{autoresttest,arat-rl, tecton}.
\Cref{tab:benchmarks} presents 
the summary statistics of data set, its benchmarks, their LOC, and number of endpoints.

For all experiments, we provide \projName with only the TypeSpec API specifications --- no pre-existing mocks, implementation sources, or external dependencies.
Although TypeSpec provides richer API specifications compared to traditional formats like OpenAPI, for the purposes of this evaluation, we generated the TypeSpec versions 
directly from the existing OpenAPI specifications via a mechanical OpenAPI-to-TypeSpec conversion and inspected to ensure that no additional data or behavior was introduced.

\projName generates a test suite for each service, writing all the test files to disk. It then starts the service under test (SUT) in a clean state and executes it on the  
generated tests. As each input is run, \projName generates appropriate mock responses and validates the behavior of the SUT. At no point is any data or network traffic sent by the SUT to actual external services --- \projName fully mocks all interactions.

All experiments were conducted on a workstation equipped with an AMD 16 core Ryzen™ 9 9950X processor, 128 GB of RAM, and coverage measurements 
are taken with the Jacoco framework~\cite{jacoco}. To avoid spurious coverage results, we configure Jacoco to exclude all test and test-support files. 
The default LLM model used for \projName test generation is deepseek-chat (DeepSeek-V3.1) in non-thinking mode. 

All tables report the median coverage values or average number of distinct 500 errors found per test generation run over $10$ runs. 
The maximum variance in coverage results across \projName runs is $3\%$ in line/branch coverage and the same set of 500 errors was 
detected on every run.

\subsection*{RQ1: Code Coverage and Error Detection}
\label{sec:topline}
The first research question (RQ1) investigates the effectiveness of \projName in terms of code coverage and error detection across different RESTful service benchmarks. For this evaluation, 
we use \tecton~\cite{tecton}, with the application-provided mocks or live services running, as the baseline for comparison against \projName. Prior work with \tecton has demonstrated its 
effectiveness in generating tests for RESTful APIs, outperforming other SotA approaches including higher code coverage and rate of  
error detection in these benchmarks.

As shown in \Cref{fig:models_comparison}, \projName consistently produces high quality test coverage, in both line and branch coverage, when compared to \tecton with application-provided 
mocks or live services. In $3$ of the benchmarks, \projName improves the coverage over \tecton with the application-provided mocking or live testing configurations. 
In the remaining $7$ benchmarks, \projName achieves comparable coverage to application-provided mocks or live services, demonstrating its robustness across different scenarios.

Although the improvement over SotA is important, an equally important objective is to meet industry-standard criteria for what would be considered ``sufficient'' or ``high-quality'' test suites. 
According to standard guidelines for code coverage, such as those recommended by Google~\cite{googletesting}, coverage above $60\%$ are considered sufficient, 
coverage above $75\%$ is regarded as high-quality, while over $90\%$ is considered exceptional. Notably, $100\%$ coverage is usually considered an anti-pattern, associated 
with low quality tests (\eg unlikely to expose actual faults), excessive CI costs, and maintenance overhead. Under this general guideline, the results in \Cref{fig:models_comparison} show 
that \projName achieves coverage levels that meet or exceed these industry standards in nine of ten benchmarks, with $3$ exceptional, $4$ high-quality, and $2$ 
sufficient coverage, demonstrating that \projName facilitates the generation of high-quality test suites.

\begin{table}
\caption{Distinct 500 Errors triggered with \tecton and application-provided mocks/live services \vs \projName generation.}
\label{tab:server_errors}
\centering
\begin{tabular}{lrr}
\hline
REST Service & \tecton & \projName \\
\hline
CWA               & 0 & 5  \\
Cat Watch         & 5 & 6 \\
Feature Service   & 9 & 13 \\
Genome Service    & 7 & 22 \\
PetClinic         & 22 & 30 \\
User Management   & 1 & 12 \\
Project Tracking  & 2 & 7 \\
Person Controller & 9 & 11 \\
Market            & 5 & 12 \\
Rental Service    & 0 & 1 \\
\hline
\end{tabular}
\end{table}

In addition to raw coverage improvements, \projName also increases the number of exposed application failures, as illustrated in \Cref{tab:server_errors}. This figure reports the 
number of distinct 500 errors encountered during testing for \tecton, with application-provided mocks or live services, \vs \projName. Each failure was binned base on the exception 
line number and de-duplicated. Representative values for each bin were used to manually verify that the error was, in fact, a real bug \emph{and} that it was a distinct 
root cause from other failures -- thus each failure represent a distinct bug and fix that would be needed.

In all benchmarks, \projName exposes more distinct 500 errors than \tecton,
finding 119 compared to 60. In the \cf{Genome} benchmark, which has the largest
number of external dependencies, the difference increases from $7$ to $22$!

A critical observation is that, even in cases where \projName did not improve code coverage metrics, such the \cf{Market} or \cf{Project Tracking} benchmarks, it still exposed 
additional server failures. Thus, \projName's benefits are not captured by coverage alone: its meaningful and diverse mock responses expose failures that prior work missed 
despite comparable coverage.

\subsection*{RQ2: Test \& Mock Generation Costs}
\label{sec:gencost}

A key consideration in evaluating \projName is the cost associated with test and mock generation, both in terms of 
generation time and monetary expense. These costs are summarized in \Cref{tab:benchmarks_cost}.

Across the ten benchmarks, mock generation took from $2$ to $54$ minutes with time depending heavily on the 
number and complexity of the API endpoints that required mock generation. The average cost per benchmark 
ranges from $\$0.01$ to $\$0.17$. 

\begin{table}
\caption{Average time (Minutes) for \projName mock generation and LLM inference cost (USD) per benchmark.}
\label{tab:benchmarks_cost}
\centering
\begin{tabular}{lrr}
\hline
REST Service & Time (Min) & Cost (USD) \\
\hline
CWA               & 3 & 0.05 \\
Cat Watch         & 15 & 0.07 \\
Feature Service   & 17 & 0.15 \\
Genome Service    & 54 & 0.17\\
PetClinic         & 35 & 0.08\\
User Management   & 32 & 0.07\\
Project Tracking  & 8 & 0.04\\
Person Controller & 20 & 0.05\\
Market            & 4 & 0.03\\
Rental Service    & 2 & 0.01\\
\hline
\end{tabular}
\end{table}

\projName was designed to focus LLMs on the constrained task of \emph{individual} value generation, not
entire response generation as prior work does.  This focus requires fewer
tokens, reduces the time required to generate a mock response, even for the more
complex benchmarks, and keeps the monetary cost low. 

Critically, as \projName stores (and fetches) the generated mocks from a cache, it is possible to 
serialize the generated mocks and store them as part of the test-suite. This feature allows 
reuse of the mocks across multiple test runs, reducing the need for repeated mock generation, and 
further lowering the effective monetary and computational costs.

\subsection*{RQ3: \projName Ablation}
\label{sec:components}
To understand the individual contributions of the successful value generation path (\Cref{sec:llm}) and the failure 
injection path (\Cref{sec:llmfail}), we analyze the number of 500 errors exposed by each component separately. 

The results shown in \Cref{tab:errors_distribution} show, for each REST service, the number of 500 errors 
exposed by the value generation alone (Value Generation column) and the additional errors exposed by the 
adding failure injection (Failure Injection column). As can be seen, both components contribute to the exposure 
of 500 errors. Interestingly, despite the conventional wisdom that failure injection is a major driver 
for exposing 500 errors in otherwise untested (or lightly tested) code, our results show that the ability 
to generate a diverse set of successful responses can be just as effective in uncovering previously 
unexplored program state and triggering errors.

\begin{table}
\caption{Features exposing each 500 error --- Value Generation reporting the number of errors exposed when 
\emph{only generating successful responses} (\Cref{sec:llm}) \vs Failure Injection reporting the number of additional errors exposed 
when \emph{injecting failures} (\Cref{sec:llmfail}).}
\label{tab:errors_distribution}
\centering
\begin{tabular}{lrr}
\hline
REST Service & Value Generation & Failure Injection \\
\hline
CWA               & 2 & 3 \\
Cat Watch         & 5 & 1 \\
Feature Service   & 3 & 10 \\
Genome Service    & 9 & 13 \\
PetClinic         & 24 & 6 \\
User Management   & 1 & 11 \\
Project Tracking  & 1 & 6 \\
Person Controller & 4 & 7 \\
Market            & 4 & 8 \\
Rental Service    & 1 & 0 \\
\hline
\end{tabular}
\end{table}

In addition, we identified a (smaller) number of failure cases where the ability to do multi-trajectory 
exploration, and mixing a combination of failure and successful responses across multiple operations, 
was critical to uncovering a failure. A key question for future work is exploring more systematic 
approaches for exploring these interactions and potentially correlating state in mock generated 
responses~\cite{chaosmusuvathi,restct,restler}.

\subsection*{RQ4: Test Suite Performance and Reliability}
\label{sec:cistuff}

Although mocking is usually done to eliminate the need to deploy and manage testing environments, it can also have significant implications for CI pipeline performance and reliability. 
\projName transforms a test run from a dependency-heavy execution, and one where we may need to block on multiple HTTP or database operations, to one that is entirely self-contained and 
all ``external operations'' are transformed into local memory lookups. As a result a single test, which may have previously taken multiple seconds or even minutes to setup and then execute, 
can now run in milliseconds entirely locally.

\Cref{tab:test_execution_time} shows the timing and test stability (flakiness) of the benchmarked REST services when using their application-provided harness compared to using \projName.
The average time (over 10 runs) in minutes taken to run the tests using the corresponding harness (application-provided or \projName). The Stable column indicates if the test suite was stable (no flakiness) 
during execution. 

\begin{table}
\caption{Comparison of test execution time and stability of test suites. Time columns indicate the average time taken to run the tests in minutes 
with application-provided testing harness and \projName. The Stable column indicates if the test suite was stable (no flakiness) during execution.}
\label{tab:test_execution_time}
\centering
\begin{tabular}{lrcrc}
\hline
& \multicolumn{2}{c}{Tecton}  & \multicolumn{2}{c}{\projName} \\
REST Service & Time (Min) & Stable & Time (Min) & Stable \\
\hline
CWA               & 1  & \textcolor{green}{$\checkmark$} & 1  & \textcolor{green}{$\checkmark$} \\
Cat Watch         & 60 & \textcolor{red}{$\times$} & 4  & \textcolor{green}{$\checkmark$} \\
Feature Service   & 3  & \textcolor{green}{$\checkmark$} & 3  & \textcolor{green}{$\checkmark$} \\
Genome Service    & 67 & \textcolor{red}{$\times$} & 12  & \textcolor{green}{$\checkmark$} \\
PetClinic         & 24 & \textcolor{green}{$\checkmark$} & 22 & \textcolor{green}{$\checkmark$} \\
User Management   & 19 & \textcolor{green}{$\checkmark$} & 19 & \textcolor{green}{$\checkmark$} \\
Project Tracking  & 3  & \textcolor{green}{$\checkmark$} & 3  & \textcolor{green}{$\checkmark$} \\
Person Controller & 17 & \textcolor{green}{$\checkmark$} & 14 & \textcolor{green}{$\checkmark$} \\
Market            & 1  & \textcolor{green}{$\checkmark$} & 1  & \textcolor{green}{$\checkmark$} \\
Rental Service    & 2  & \textcolor{red}{$\times$} & 1  & \textcolor{green}{$\checkmark$} \\
\hline
\end{tabular}
\end{table}

The results in \Cref{tab:test_execution_time} demonstrate that \projName significantly improves test stability and execution time for benchmarks with external dependencies. 
In particular, benchmarks that rely on external services, such as hosted Git repositories or remote APIs, benefit the most from \projName's ability to replace these dependencies with local mocks 
and see (sequential) runtimes drop by up to $15\times$. Further, by replacing calls to live external services (and forcing interleavings) \projName is able to eliminate 
the flakey behavior seen in several cases (such as the \cf{Catwatch} benchmark which makes calls to \url{github.com} during test execution). 

\subsection{Threats to Validity}
\label{sec:threats}
We discuss four threats to the validity of our results. The first concerns the construct validity of our baseline; 
the remaining three concern the extent to which our findings generalize beyond the studied 
services, specification language, and evaluation setting.\\

\noindent
\emph{Fidelity of the baseline application harnesses and mocks.}
Our comparisons in RQ1, RQ3, and RQ4 use the harness and/or mocks provided as part of the benchmark application as the baseline. 
Our results depend on the quality of these baseline mocks and the extent to which they effectively test the behavior of the application.
Although we have selected benchmarks with existing mocks and specifications, which are used broadly in the research literature~\cite{prab}, the 
quality and completeness of the provided harnesses, including mocks and test inputs, has a direct influence on our comparative results.\\

\noindent
\emph{Benchmark selection and representativeness.}
Our subjects are ten Java Spring Boot services drawn from an established public REST benchmark, 
filtered by the criteria described above and further restricted to services with an existing OpenAPI 
specification. This selection introduces possible biases including the ecosystem homogeneity, application scale, 
requirement specification availability criteria, and total number of benchmarks. As with any benchmark study, our results are subject to 
the assumption that that these applications are representative of the broader population of Java Spring Boot services.\\

\noindent
\emph{Dependence on TypeSpec.}
\projName consumes TypeSpec specifications~\cite{typespec}. For this evaluation, every specification was obtained by 
mechanical OpenAPI-to-TypeSpec conversion, and were inspected to ensure that the conversion did not introduce any 
additional constraints or behavioral information. This design isolates the contribution of \projName from the 
expressiveness of the specification language. However, this conversion may not accurately represent the use 
of TypeSpec in practice. Further, we do not attempt to evaluate how our results transfer to other specification 
languages (\eg Smithy~\cite{smithy}). Thus, our findings are specific to the OpenAPI-to-TypeSpec conversion 
and may not generalize to other specification languages or native TypeSpec usage.\\

\noindent
\emph{Error validity and transfer of false-positive rates.}
We use distinct HTTP 500 responses as our failure oracle, binned by the throwing source location and manually 
confirmed to be genuine, distinct defects. This methodology has three limitations. 
First, manual classification by the authors is subjective and subject to human error. Additionally, \projName 
generates \emph{spec-anticipatory} responses: values that satisfy the specification but that the current dependency 
implementation may never return, including value combinations that are individually valid but jointly implausible 
(\Cref{sec:llm}). This may lead to failures that are flagged as defects by our oracle but that would never occur 
in practice with the current dependency implementation and would be treated as \emph{false positives}. In our
experimental results, while unrealistic values were generated, all 500 error bins contained at least one \emph{true positive} 
witness -- indicating a genuine defect. Further, we claim that spec-anticipatory responses, even if currently infeasible, represent 
true latent defects that could manifest in future versions of the dependency and should be treated as such.

\section{Related Work}
\label{sec:relatedwork}
Given the importance of, and challenges around, mocking in software testing there is an extensive history 
of research in the area, some of which has been cited throughout the prior sections. In this section we 
focus on key results and high level trends.\\

\noindent
\emph{Practical Mocking Challenges:}
Empirical studies of mocking practices in software development have highlight the range of challenges 
developers face when creating and maintaining mocks~\cite{mockingstudy,mockingstudy2}. 
These range from selecting which behaviors to model, the implementation effort required to create accurate mocks, 
and the ongoing maintenance burden as the system evolves. The result is that developers report 
skipping certain behaviors in their mocks, at the cost of test-suite quality, and focusing instead 
only on the most common or straightforward cases.\\

\noindent
\emph{Flakey Tests and Chaos Testing:}
A major challenge when testing is the non-deterministic behavior of external services which, if 
not controlled or properly mocked, can lead to flaky tests that undermine confidence in the test suite. 
An empirical study by \cite{flakeytestexperience} investigates the developer experience with flaky tests, 
highlighting the challenges they pose for maintaining reliable test suites and continuous integration pipelines.

A related issue the is difficulty of testing failure scenarios -- such as network delays or odd interleavings 
of operations. These are often overlooked by developers when writing tests, leading to gaps in test coverage 
and potential reliability issues, and are difficult to reproduce and test effectively. Techniques 
such as those proposed in \cite{coyote,chaosmusuvathi,chaosmonkey} aim to address these challenges by 
systematically introducing failures and operations reorderings in a controlled manner to explicitly 
test these scenarios.\\

\noindent
\emph{Automated Mock Generation}
Given these challenges, automated mock generation is a long-standing topic of research. Given the history 
there have been numerous approaches for generating the mock data/responses. These range from symbolic 
solving~\cite{pexmoles}, to learning from existing production data~\cite{martinmockgen}, and applying 
search based fuzzing~\cite{fuzzmock,stubcoder}, and of course using LLMs for mock generation~\cite{ning2026executable}. 

Interestingly, Ning \etal~\cite{ning2026executable} show the direct use of LLMs for generating mock data is unsatisfactory as ``...LLMs and 
coding agents synthesize mocks merely to satisfy tests...'', a similar issue of writing tests ``merely to cover code 
boost metrics'' is also observed~\cite{llmbuggyfollow,teststruggle} when using LLMs to generate input values as well.   
Seran \etal~\cite{fuzzmock} note multi-level schema extraction as a key an important future work item. Both of which are key features that 
\projName directly addresses these challenges by incorporating multi-level schema extraction and focused LLM 
operations in the mock generation process.\\

\noindent
\emph{Agentic Tool Synthesis/Simulation:}
The topic of tool simulation and synthesis is closely related to automated mock generation, as both involve 
creating a controlled set of responses for simulating external systems for testing and evaluation 
purposes. ToolEmu, ToolAlpaca~\cite{toolemu,toolalpaca} are examples of such frameworks, and use an LLM 
to simulate responses to tool uses (API calls) based on the arguments that an agent passes during its operation.

\section{Conclusion} %Onward!
\label{sec:conclusion}

Testing service-based applications requires exploring both their inputs and behaviours of their dependencies.
This work introduced \projName, which automates mock generation by combining structural decomposition, focused 
LLM generation of component values, and systematic recombination. This design allows 
\projName to construct diverse mock responses, and in combination with failure-injection and multi-trajectory testing, 
allows systematic exploration of both normal and failure scenarios.

Across $10$ RESTful service benchmarks, \projName improves coverage on three and achieves comparable coverage on the
remaining seven, while exposing more distinct server errors on every benchmark than testing with developer-written
mocks or live services. This demonstrates that response diversity can improve fault detection and that coverage
is a weak indicator of testing quality. These results, and the possibilities for improved test-suite performance, also demonstrate 
that \projName, and the approach it takes, represent a significant step towards more robust and comprehensive testing of service-based applications.

\section*{AI Use}
The production and analysis of all results in this paper were directly performed by the authors. The content of this paper is human-authored with AI assistance 
for figure generation, reviews for data transcription errors, formatting issues, and proofing suggestions to improve the manuscript.

\section*{Data-Availability Statement}
\label{sec:datavailibility}
All code, data, and benchmarks used in this study are in active development, are publicly available, and open source licensed at 
\url{https://github.com/BosqueLanguage/BosqueCore} and \url{https://github.com/BosqueLanguage/BSQON}. 

An archival reproduction package is available to facilitate independent use of our results and is 
located at \url{https://zenodo.org/records/24181222}. This data includes a runnable version of the tool, the parameters 
used for the results in this paper, and archived output test suites. 

%\begin{acks}
%The authors would like to thank... \todo{this}
%\end{acks}

\bibliographystyle{ACM-Reference-Format}
\bibliography{bibfile}

%\section*{Appendix}
%\label{sec:appendix}

\end{document}